\documentclass[aps,showpacs,showkeys,preprintnumbers,amsmath,amssymb]{revtex4}
\usepackage{graphicx}
\usepackage{slashed}
\begin{document}

\title{
Calculation of $1/m^{2}_{b}$ corrections to $\langle\Lambda_{b}{(v,s)}|\bar{b}\gamma^{\lambda}\gamma_{5}b|\Lambda_{b}(v,s)\rangle$ for polarized $\Lambda_{b}$ in the Bethe-Salpeter equation approach}

\author{ L. Zhang$^{1}$\footnote[1]{
200931220004@mail.bnu.edu.cn}, X.-H. Guo$^{1}$\footnote[2]{ Corresponding author. xhguo@bnu.edu.cn}}
\affiliation{$^{1}$ College of Nuclear Science and Technology,
Beijing Normal University, Beijing
100875, People's Republic of China }


\begin{abstract}

 The heavy baryon $\Lambda_{Q}~(Q=b$ or ~$c)$ can be regarded as composed of a heavy quark and a scalar light diquark which has good spin and isospin quantum numbers. In this picture we establish the Bethe-Salpeter (BS) equation for $\Lambda_{Q}$ to second order in the $1/m_{Q}$ expansion. With the kernel containing both the scalar confinement and the one-gluon-exchange terms we solve the BS equation numerically. The value of the spin-dependant form factor for the matrix element $\langle\Lambda_{b}{(v,s)}|\bar{b}\gamma^{\lambda}\gamma_{5}b|\Lambda_{b}(v,s)\rangle$, $\epsilon_{b}$, which is non-zero at order $1/m^{2}_{b}$, is obtained numerically from our model.

\end{abstract}




\maketitle

\section*{I. Introduction}

In the past decades there has been much progress in heavy flavor physics due to the discovery of the new flavor and spin symmetries $SU(2)_{f}\times SU(2)_{s}$ in the heavy quark limit and the establishment of the heavy quark effective theory (HQET)~\cite{isg89}. The Large Hadron Collider (LHC) will provide much more data for heavy hadrons, and hence it will be able to test the standard model (SM) more accurately. One may expect more precise measurement of Cabibbo-Kobayashi-Maskawa matrix elements such as $V_{ub}$ in the near future.

There have been extensive studies in literature on inclusive semileptonic decays of bottom hadrons, $H_{b}\rightarrow Xe\bar{\nu}_{e}$ ($X$ represents all hadrons in the final states)~\cite{Chay90}-\cite{Ura05}, especially since the establishment of HQET. These studies include corrections to the leading order results both from perturbative QCD ($\alpha_{s}(m_{b})$) terms and from nonperturbative terms which are suppressed by powers of $m_{b}$. It was pointed out that there are no $1/m_{b}$ corrections to leading order in the $1/m_{b}$ expansion for the differential decay widths of inclusive semileptonic decays of bottom hadrons, $d\Gamma/dq^{2}dE_{e}$, where $q$ is the total momentum of the electron and the neutrino and $E_{e}$ is the electron energy in the $\Lambda_{b}$'s rest frame ~\cite{Chay90}. Bigi et al. studied ~$1/m^{2}_{b}$ corrections to the decay width $d\Gamma/dE_{e}$~\cite{Big93}. Manohar and Wise analyzed extensively $1/m^{2}_{b}$ corrections to $d\Gamma/dq^{2}dE_{e}$ for unpolarized bottom hadron $H_{b}$ and for polarized $\Lambda_{b}$~\cite{Man94}. In recent years, theoretical calculations for the inclusive semileptonic decay widths and for the moments of inclusive observables  have been carried out to order $1/m^{3}_{b}$ and $\alpha^{2}_{s}\beta_{0} ~(\beta_{0}=11-2n_{f}/3,$~$n_{f}$ is the number of quark flavors)~\cite{Fal98}-\cite{Ura05}.

At leading order in $\alpha_{s}(m_{b}$) the polarized differential semileptonic decay rate for $\Lambda_{b}\rightarrow X_{u}e\bar{\nu}_{e}$~(where $X_{u}$ represents all hadrons containing an up quark) is
\begin{eqnarray}
\frac{d\Gamma}{dq^{2}dE_{e}dE_{\nu}d\cos\theta}&=&\frac{|V_{jb}|^{2}G^{2}_{F}}{2\pi^{3}}\Big[W_{1}q^{2}+W_{2}\Big(2E_{e}E_{\nu}-\frac{1}{2}q^{2}\Big)+W_{3}q^{2}(E_{e}-E_{\nu})\Big]\nonumber\\
&&+\frac{|V_{jb}|^{2}G^{2}_{F}}{4\pi^{3}}\cos\theta\Bigg\{\Big[G_{1}q^{2}+G_{2}\bigg(2E_{e}E_{\nu}-\frac{1}{2}q^{2}\bigg)+G_{3}q^{2}|(E_{e}-E_{\nu})|\Big]\bigg(E_{e}+E_{\nu}-\frac{q^{2}}{2E_{e}}\bigg)\nonumber\\
&&+G_{6}(q^{2}-4E_{e}E_{\nu})-G_{8}q^{2}-G_{9}q^{2}\bigg(E_{e}-E_{\nu}+\frac{q^{2}}{2E_{e}}\bigg)\Bigg\},
\end{eqnarray}
\noindent where the electron mass is neglected, $j=u,c$,~$\theta$ is the angle between the electron three-momentum and the spin vector of $\Lambda_{b}$ in the $\Lambda_{b}$'s rest frame, $E_{\nu}$ is the neutrino energy in the rest frame of $\Lambda_{b}$, $q^{2}=(p_{e}+p_{\nu})^{2}$ is the invariant mass of the lepton pair, the kinematic variables are to be integrated over the region ~$q^{2}\leq 4E_{e}E_{\nu}$, $W_{1}$, $W_{2}$, and $W_{3}$ are the form factors in spin-independent hadronic tensor $W^{\mu\nu}$, $G_{1}$, $G_{2}$, $G_{3}$, $G_{6}$, $G_{8}$, and $G_{9}$ are the form factors in spin-dependent hadronic tensor ~$W^{\mu\nu}_{S}$.~$W^{\mu\nu}$ and~$W^{\mu\nu}_{S}$ have the following forms \cite{Man94}:
\begin{eqnarray}
W^{\mu\nu}=-g^{\mu\nu}W_{1}+v^{\mu}v^{\nu}W_{2}-i\epsilon^{\mu\nu\alpha\beta}v_{\alpha}q_{\beta}W_{3}+q^{\mu}q^{\nu}W_{4}+(q^{\mu}v^{\nu}+q^{\nu}v^{\mu})W_{5},
\end{eqnarray}
\noindent and
\begin{eqnarray}
W^{\mu\nu}_{S}&=&-q\cdot s[-g^{\mu\nu}G_{1}+v^{\mu}v^{\nu}G_{2}-i\epsilon^{\mu\nu\alpha\beta}v_{\alpha}q_{\beta}G_{3}+q^{\mu}q^{\nu}G_{4}+(q^{\mu}v^{\nu}+q^{\nu}v^{\mu})G_{5}]\nonumber\\
&&+(s^{\mu}v^{\nu}+s^{\nu}v^{\mu})G_{6}+(s^{\mu}q^{\nu}+s^{\nu}q^{\mu})G_{7}+i\epsilon^{\mu\nu\alpha\beta}v_{\alpha}s_{\beta}G_{8}+i\epsilon^{\mu\nu\alpha\beta}q_{\alpha}s_{\beta}G_{9}),
\end{eqnarray}
\noindent where~$s^{\mu}$ is the spin vector of $\Lambda_{b}$ and $v^{\mu}$ is the velocity of $\Lambda_{b}$.

When the electron mass is ignored, the form factors ~$W_{4}$,~$W_{5}$,~$G_{4}$,~$G_{5}$, and~$G_{7}$ in Eqs.(2)(3) do not contribute to the differential inclusive semleptomic decay rate of polarized~$\Lambda_{b}$~\cite{Man94}. The expressions for other form factors,~$W_{1}$, $W_{2}$, $W_{3}$, $G_{1}$, $G_{2}$, $G_{3}$,~$G_{6}$,~$G_{8}$, and~$G_{9}$, were obtained to second order in the $1/m_{b}$ expansion in Ref.~\cite{Man94}. It was shown that $1/m^{2}_{b}$ corrections were characterized by two parameters, $\mu^{2}_{\pi}$ and $\epsilon_{b}$, for polarized $\Lambda_{b}$ decays~(the spin energy is zero for $\Lambda_{b}$). $\mu^{2}_{\pi}$ is the kinetic energy which is defined as
\begin{eqnarray}
\mu^{2}_{\pi}=-\frac{\langle\Lambda_{b}|\bar{h}_{v}(iD_{\bot})^{2}h_{v}|\Lambda_{b}\rangle}{2m_{b}},
\end{eqnarray}
\noindent where $h_{v}$ denotes the field of $b$ quark in HQET, $D^{\mu}_{\bot}=D^{\mu}-v^{\mu}v\cdot D$ with $D^{\mu}$ being the covariant derivative. The value of $\mu^{2}_{\pi}$ was calculated for the first time in Ref.~\cite{Guo07}. $\epsilon_{b}$ is defined as
\begin{eqnarray}
\langle\Lambda_{b}(v,s)|\bar{b}\gamma^{\lambda}\gamma_{5}b|\Lambda_{b}(v,s)\rangle&=&(1+\epsilon_{b})\bar{u}_{\Lambda_{b}}(v,s)\gamma^{\lambda}\gamma_{5}u_{\Lambda_{b}}(v,s)\nonumber\\
&=&(1+\epsilon_{b})s^{\lambda},
\end{eqnarray}
\noindent where $u_{\Lambda_{b}}(v,s)$ is the Dirac spinor of $\Lambda_{b}$ with helicity $s$. $\epsilon_{b}$ is equal to zero at leading order in the $1/m_{b}$ expansion due to the heavy quark symmetry. It is also zero at the first order in $1/m_{b}$ expansion because of the current conservation~\cite{Luk90}. It is the aim of the present paper to calculate the value of $\epsilon_{b}$ at order $1/m^{2}_{b}$. This will be important for the precise measurement of $V_{ub}$ in polarized $\Lambda_{b}$ decays.

When the quark mass is very heavy compared with the QCD scale $\Lambda_{QCD}$, the light degrees of freedom in a heavy baryon $\Lambda_{Q}$ ($Q=b$ or $c$) become blind to the flavor and spin quantum numbers of the heavy quark because of the $SU(2)_{f}\times SU(2)_{s}$ symmetries. Therefore, the angular momentum and flavor quantum numbers of the light degrees of freedom become good quantum numbers which can be used to classify heavy baryons, and $\Lambda_{Q}$ corresponds to the state in which the angular momentum of the light degrees of freedom is zero. So it is natural to regard the heavy baryon as composed of one heavy quark and a light diquark. When $1/m_{Q}$ corrections are taken into account, since the isospin of $\Lambda_{Q}$ is zero, the isospin of the light degrees of freedom is also zero. Therefore, the spin of the light degrees of freedom should also be zero in order to guarantee that the total wave function of $\Lambda_{Q}$ is anti-symmetric. Hence the spin and isospin of the light degrees of freedom are still fixed even when $1/m_{Q}$ corrections are taken in account. Therefore, we still treat $\Lambda_{Q}$ as composed of a heavy quark and a light diquark.

Based on the above picture, the three-body system is simplified to a two-body system. We will establish the Bethe-Salpeter (BS) equation for $\Lambda_{Q}$ in this picture to order $1/m^{2}_{Q}$. Then, we will solve this equation numerically by assuming that the kernel contains the scalar confinement and the one-gluon-exchange terms. Furthermore, we will express $\epsilon_{b}$ in terms of the BS amplitude of $\Lambda_{b}$ in order to calculate the value of $\epsilon_{b}$ to second order in the $1/m_{b}$ expansion.

The remainder of this paper is organized as follows. In Sec. II we will establish the BS equation for $\Lambda_{Q}$ to second order in the $1/m_{Q}$ expansion and discuss the form of its kernel. Then, we will solve the equation numerically. In Sec. III we will express $\epsilon_{b}$ in terms of the BS amplitude of $\Lambda_{b}$ and calculate the value of $\epsilon_{b}$ for different model parameters. Finally, Sec. IV is reserved for summary and discussion.

\section*{II. Bethe-Salpeter equation for~$\Lambda_{Q}$ to second order in the~$1/m_{Q}$ expansion}

As we discussed in Introduction, $\Lambda_{Q}$ is regarded as a bound state of a heavy quark and a light scalar diquark. Hence we can define the BS amplitude of $\Lambda_{Q}$ as the following:
\begin{eqnarray}
\chi(x_{1},x_{2},P)=\langle 0|T\psi(x_{1})\varphi(x_{2})|\Lambda_{Q}(P)\rangle,
\end{eqnarray}
\noindent where $\psi(x_{1})$ and $\varphi(x_{2})$ are the field operators of the heavy quark at position $x_{1}$ and the light scalar diquark at position $x_{2}$, respectively, $P=m_{\Lambda_{Q}}v$ is the momentum of $\Lambda_{Q}$, and $v$ is its velocity. Let $m_{Q}$ and $m_{D}$ represent the masses of the heavy quark and the light diquark in the baryon,  $\lambda_{1}=\frac{m_{Q}}{m_{Q}+m_{D}}$, $\lambda_{2}=\frac{m_{D}}{m_{Q}+m_{D}}$, and $p$ represents the relative momentum of the two constituents. Then, the BS amplitude in momentum space is defined as
\begin{eqnarray}
\chi(x_{1},x_{2},P)=e^{iPX}\int\frac{d^{4}p}{(2\pi)^{4}}e^{ipx}\chi_{P}(p),
\end{eqnarray}
\noindent where $X=\lambda_{1}x_{1}+\lambda_{2}x_{2}$ is the coordinate of the center of mass and $x=x_{1}-x_{2}$.

It is straightforward to prove that the BS equation for $\Lambda_{Q}$ has the following form in momentum space:
\begin{eqnarray}
\chi_{P}(p)=S_{F}(p_{1})\int\frac{d^{4}q}{(2\pi)^{4}}K(P,p,q)\chi_{P}(q)S_{D}(p_{2}),
\end{eqnarray}
\noindent where $p_{1}=\lambda_{1}P+p$, $p_{2}=-\lambda_{2}P+p$ are the momenta of the heavy quark and the light scalar diquark, respectively, $K(P,p,q)$ is the kernel which is defined as the sum of two particle irreducible diagrams, $S_{F}(p_{1})$ and $S_{D}(p_{2})$ are propagators of the heavy quark with momentum $p_{1}$ and the light diquark with momentum $p_{2}$,
\begin{eqnarray}
S_{F}(p_{1})=\frac{i}{\slashed{p}_{1}-m_{Q}+i\varepsilon},
\end{eqnarray}
\begin{eqnarray}
S_{D}(p_{2})=\frac{1}{p_{2}^{2}-m_{D}^2+i\varepsilon}.
\end{eqnarray}

In order to solve the BS equation for $\Lambda_{Q}$ to second order in the $1/m_{Q}$ expansion, we rewrite Eq.~(8) as the following:
\begin{eqnarray}
\chi_{0P}(p)+\frac{1}{m_{Q}}\chi_{1P}(p)+\frac{1}{m_{Q}^{2}}\chi_{2P}(p)&=&\bigg(S_{0F}(p_{1})+\frac{1}{m_{Q}}S_{1F}(p_{1})+\frac{1}{m_{Q}^{2}}S_{2F}(p_{1})\bigg)\int\frac{d^{4}q}{(2\pi)^{4}}\bigg(K_{0}(P,p,q)\nonumber\\
&&+\frac{1}{m_{Q}}K_{1}(P,p,q)+\frac{1}{m_{Q}^{2}}K_{2}(P,p,q)\bigg)\times\Big(\chi_{0P}(q)+\frac{1}{m_{Q}}\chi_{1P}(q)\nonumber\\
&&+\frac{1}{m_{Q}^{2}}\chi_{2P}(q)\Big)S_{D}(p_{2}),
\end{eqnarray}
\noindent where we have expanded the BS amplitude, the propagator of the heavy quark, and the kernel to $1/m^{2}_{Q}$.

It is easy to show that $S_{D}$ remains unchanged in the $1/m_{Q}$ expansion. Then, by comparing the two sides of Eq.~(11) at each order in $1/m_{Q}$, we have following equations:

\noindent to leading order in the ~$1/m_{Q}$ expansion:
\begin{eqnarray}
\chi_{0P}(p)=S_{0F}(p_{1})\int\frac{d^{4}q}{(2\pi)^4}K_{0}(P,p,q)\chi_{0P}(q)S_{D}(p_{2}),
\end{eqnarray}
to first order in the~$1/m_{Q}$ expansion:
\begin{eqnarray}
\chi_{1P}(p)&=&S_{1F}(p_{1})\int\frac{d^{4}q}{(2\pi)^{4}}K_{0}(P,p,q)\chi_{0P}(q)S_{D}(p_{2})+S_{0F}(p_{1})\int\frac{d^{4}q}{(2\pi)^{4}}K_{1}(P,p,q)\chi_{0P}(q)S_{D}(p_{2})\nonumber\\
&&+S_{0F}(p_{1})\int\frac{d^{4}q}{(2\pi)^{4}}K_{0}(P,p,q)\chi_{1P}(q)S_{D}(p_{2}),
\end{eqnarray}
to second order in the~$1/m_{Q}$ expansion:
\begin{eqnarray}
\chi_{2P}(p)&=&S_{2F}(p_{1})\int\frac{d^{4}q}{(2\pi)^{4}}K_{0}(P,p,q)\chi_{0P}(q)S_{D}(p_{2})+S_{1F}(p_{1})\int\frac{d^{4}q}{(2\pi)^{4}}K_{1}(P,p,q)\chi_{0P}(q)S_{D}(p_{2})\nonumber\\
&&+S_{1F}(p_{1})\int\frac{d^{4}q}{(2\pi)^{4}}K_{0}(P,p,q)\chi_{1P}(q)S_{D}(p_{2})+S_{0F}(p_{1})\int\frac{d^{4}q}{(2\pi)^{4}}K_{2}(P,p,q)\chi_{0P}(q)S_{D}(p_{2})\nonumber\\
&&+S_{0F}(p_{1})\int\frac{d^{4}q}{(2\pi)^{4}}K_{1}(P,p,q)\chi_{1P}(q)S_{D}(p_{2})+S_{0F}(p_{1})\int\frac{d^{4}q}{(2\pi)^{4}}K_{0}(P,p,q)\chi_{2P}(q)S_{D}(p_{2}).
\end{eqnarray}

In the following we will use the variables $p_{l}=v\cdot p-\lambda_{2}m_{\Lambda_{Q}}$, $p_{t}=p-(v\cdot p)v$. Then, in leading order, first order and second order in the $1/m_{Q}$ expansion, we have the following equations for the heavy quark propagator using the relation $p_{1}=\lambda_{1}P+p$~\cite{Guo96}~\cite{Guo00}:
\begin{eqnarray}
S_{0F}(p_{1})=\frac{i(1+\slashed{v})}{2(p_{l}+E_{0}+m_{D}+i\varepsilon)},
\end{eqnarray}
\begin{eqnarray}
S_{1F}(p_{1})=-\frac{i(1-\slashed{v})}{4}+\frac{i\slashed{p}_{t}}{2(p_{l}+E_{0}+m_{D}+i\varepsilon)}-\frac{i(2E_{1}-p^{2}_{t})(1+\slashed{v})}{4(p_{l}+E_{0}+m_{D}+i\varepsilon)^{2}},
\end{eqnarray}
\begin{eqnarray}
S_{2F}(p_{1})&=&\frac{i(p_{l}+E_{0}+m_{D})(1+\slashed{v})}{8}-\frac{i\slashed{p}_{t}}{4}-\frac{ip^{2}_{t}}{4(p_{l}+E_{0}+m_{D}+i\varepsilon)}\nonumber\\
&&-\frac{2iE_{2}(1+\slashed{v})+i(2E_{1}-p^{2}_{t})\slashed{p}_{t}}{4(p_{l}+E_{0}+m_{D}+i\varepsilon)^{2}}+\frac{i(2E_{1}-p^{2}_{t})^{2}(1+\slashed{v})}{8(p_{l}+E_{0}+m_{D}+i\varepsilon)^{3}},
\end{eqnarray}
\noindent where $E_{0}$, $E_{1}$, and $E_{2}$ represent the binding energies, which satisfy the following relation:
\begin{eqnarray}
m_{\Lambda_{Q}}=m_{Q}+m_{D}+E_{0}+\frac{1}{m_{Q}}E_{1}+\frac{1}{m_{Q}^{2}}E_{2}.
\end{eqnarray}

$S_{D}(p_{2})$ remains unchanged in the $1/m_{Q}$ expansion, and we rewrite it as the following using the relation $p_{2}=-\lambda_{2}P+p$:
\begin{eqnarray}
S_{D}(p_{2})=\frac{i}{p^{2}_{l}-W^{2}_{p}+i\varepsilon},
\end{eqnarray}
\noindent where $W^{2}_{p}=\sqrt{p^{2}_{t}+m^{2}_{D}}$.

\begin{center}
\begin{figure}[htbp]
\includegraphics[height=6cm]{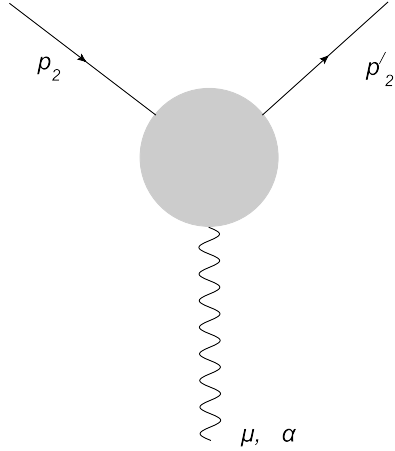}
\caption{~Diquark-gluon-diquark vetex, $\mu$ and $\alpha$ are the Lorentz and color indices of the gluon, respectively.}
\end{figure}
\end{center}

We assume the kernel has the following form:
\begin{eqnarray}
-iK_{0}=1\otimes 1V_{1}+v^{\mu}\otimes(p_{2}+p'_{2})^{\mu}V_{2},
\end{eqnarray}
\begin{eqnarray}
-iK_{1}=1\otimes 1V_{3}+\gamma^{\mu}\otimes(p_{2}+p'_{2})^{\mu}V_{4},
\end{eqnarray}
\begin{eqnarray}
-iK_{2}=1\otimes 1V_{5}+\gamma^{\mu}\otimes(p_{2}+p'_{2})^{\mu}V_{6},
\end{eqnarray}
\noindent where the first terms on the right hand sides of Eqs.~(20), (21), and (22) arise from the scalar confinement and the second ones are from the one-gluon-exchange diagram, $p_{2}$ and $p'_{2}$ are the momenta of the light diquark attached to the gluon. $v^{\mu}$ and $\gamma^{\mu}$ in Eqs. (20)-(22) are from the vertex of the heavy quark and the gluon. It is noted that $\gamma^{\mu}$ in Eq.~(20) becomes $v^{\mu}$ because of heavy quark symmetry (see FIG.~1). The forms of $V_{1}$ and $V_{2}$ were given before~\cite{Guo96}. In this work, we assume that $V_{3}$ and $V_{4}$, $V_{5}$ and $V_{6}$ have the same forms as $V_{1}$ and $V_{2}$, respectively. However, $V_{3}$ and $V_{4}$, $V_{5}$ and $V_{6}$ are suppressed with respect to $V_{1}$ and $V_{2}$ by order $\Lambda_{QCD}/m_{Q}$ and $\Lambda^{2}_{QCD}/m^{2}_{Q}$, respectively. Note that $v^{\mu}$ in Eq.~(20) is replaced by $\gamma^{\mu}$ in Eqs.~(21) and (22) since there is no heavy quark spin symmetry when we include $1/m_{Q}$ corrections.

Combining Eqs.~(12)-(22), we have the BS equations for $\Lambda_{Q}$ in leading order, first order, and second order in the $1/m_{Q}$ expansion:
\begin{eqnarray}
\chi_{0P}(p)
&=&\frac{i(1+\slashed{v})}{2(p_{l}+m_{D}+E_{0}+i\varepsilon)}\int\frac{d^{4}q}{(2\pi)^{4}}i\big[1\otimes1V_{1}+v_{\mu}\otimes(p_{2}+p'_{2})^{\mu}V_{2}\big]\chi_{0P}(q)\frac{i}{p^{2}_{l}-W^{2}_{P}+i\varepsilon},
\end{eqnarray}
\begin{eqnarray}
\chi_{1P}(p)&=&\frac{i(1+\slashed{v})}{2(p_{l}+m_{D}+E_{0}+i\varepsilon)}\int\frac{d^{4}q}{(2\pi)^{4}}i[1\otimes1V_{3}+\gamma_{\mu}\otimes(p_{2}+p'_{2})^{\mu}V_{4}]\chi_{0P}(q)\frac{i}{p^{2}_{l}-W^{2}_{P}+i\varepsilon}\nonumber\\
&&+\frac{i(1+\slashed{v})}{2(p_{l}+m_{D}+E_{0}+i\varepsilon)}\int\frac{d^{4}q}{(2\pi)^{4}}i[1\otimes1V_{1}+v_{\mu}\otimes(p_{2}+p'_{2})^{\mu}V_{2}]\chi_{1P}(q)\frac{i}{p^{2}_{l}-W^{2}_{P}+i\varepsilon}\nonumber\\
&&+i\bigg[\frac{(-2E_{1}+p^{2}_{t})(1+\slashed{v})}{4(p_{l}+m_{D}+E_{0}+i\varepsilon)^{2}}-\frac{1-\slashed{v}}{4}+\frac{\slashed{p}_{t}}{2(p_{l}+m_{D}+E_{0}+i\varepsilon)}\bigg]\nonumber\\
&&~~~~~~~~~~~\times i\int\frac{d^{4}q}{(2\pi)^{4}}[1\otimes1V_{1}+v_{\mu}\otimes(p_{2}+p'_{2})^{\mu}V_{2}]\chi_{0P}(q)\frac{i}{p^{2}_{l}-W^{2}_{P}+i\varepsilon},
\end{eqnarray}
\begin{eqnarray}
\chi_{2P}(p)&=&\frac{i(1+\slashed{v})}{2(p_{l}+m_{D}+E_{0}+i\varepsilon)}\int\frac{d^{4}q}{(2\pi)^{4}}i[1\otimes1V_{1}+v_{\mu}\otimes(p_{2}+p'_{2})^{\mu}V_{2}]\chi_{2P}(q)\frac{i}{p^{2}_{l}-W^{2}_{P}+i\varepsilon}\nonumber\\
&&+\frac{i(1+\slashed{v})}{2(p_{l}+m_{D}+E_{0}+i\varepsilon)}\int\frac{d^{4}q}{(2\pi)^{4}}i[1\otimes1V_{3}+\gamma_{\mu}\otimes(p_{2}+p'_{2})^{\mu}V_{4}]\chi_{1P}(q)\frac{i}{p^{2}_{l}-W^{2}_{P}+i\varepsilon}\nonumber\\
&&+\frac{i(1+\slashed{v})}{2(p_{l}+m_{D}+E_{0}+i\varepsilon)}\int\frac{d^{4}q}{(2\pi)^{4}}i[1\otimes1V_{5}+\gamma_{\mu}\otimes(p_{2}+p'_{2})^{\mu}V_{6}]\chi_{0P}(q)\frac{i}{p^{2}_{l}-W^{2}_{P}+i\varepsilon}\nonumber\\
&&+i\bigg[\frac{(-2E_{1}+p^{2}_{t})(1+\slashed{v})}{4(p_{l}+m_{D}+E_{0}+i\varepsilon)^{2}}-\frac{1-\slashed{v}}{4}+\frac{\slashed{p}_{t}}{2(p_{l}+m_{D}+E_{0}+i\varepsilon)}\bigg]\nonumber\\
&&~~~~~~~~~\times\int\frac{d^{4}q}{(2\pi)^{4}}i[1\otimes1V_{1}+v_{\mu}\otimes(p_{2}+p'_{2})^{\mu}V_{2}]\chi_{1P}(q)\frac{i}{p^{2}_{l}-W^{2}_{P}+i\varepsilon}\nonumber\\
&&+i\bigg[\frac{(-2E_{1}+p^{2}_{t})(1+\slashed{v})}{4(p_{l}+m_{D}+E_{0}+i\varepsilon)^{2}}-\frac{1-\slashed{v}}{4}+\frac{\slashed{p}_{t}}{2(p_{l}+m_{D}+E_{0}+i\varepsilon)}\bigg]\nonumber\\
&&~~~~~~~~~\times\int\frac{d^{4}q}{(2\pi)^{4}}i[1\otimes1V_{3}+\gamma_{\mu}\otimes(p_{2}+p'_{2})^{\mu}V_{4}]\chi_{0P}(q)\frac{i}{p^{2}_{l}-W^{2}_{P}+i\varepsilon}\nonumber\\
&&+i\bigg[\frac{(p_{l}+m_{D}+E_{0})(1+\slashed{v})}{8}-\frac{\slashed{p}_{t}}{4}+\frac{-p^{2}_{t}}{4(p_{l}+m_{D}+E_{0}+i\varepsilon)}-\frac{2E_{2}(1+\slashed{v})}{4(p_{l}+m_{D}+E_{0}+i\varepsilon)^{2}}\nonumber\\
&&~~-\frac{(2E_{1}-p^{2}_{t})\slashed{p}_{t}}{4(p_{l}+m_{D}+E_{0}+i\varepsilon)^{2}}+\frac{(2E_{1}-p^{2}_{t})^{2}(1+\slashed{v})}{8(p_{l}+m_{D}+E_{0}+i\varepsilon)^{3}}\bigg]\int\frac{d^{4}q}{(2\pi)^{4}}i[1\otimes1V_{1}+v_{\mu}\otimes(p_{2}+p'_{2})^{\mu}V_{2}]\nonumber\\
&&~~~~~~~~~~\times\chi_{0P}(q)\frac{i}{p^{2}_{l}-W^{2}_{P}+i\varepsilon}.
\end{eqnarray}

It is noted that Eqs.~(23) and (24) are just the equations obtained before in the heavy quark limit~\cite{Guo96}, and to first order in the $1/m_{Q}$ expansion~\cite{Guo00}.

In general, $\chi_{P}(p)$ can be expanded as
\begin{equation}
\chi_{P}(p)=(A+B\slashed{v}+C\slashed{p}+D\slashed{v}\slashed{p})u_{\Lambda_{Q}}(v,s),
\end{equation}
\noindent where $A$, $B$, $C$, $D$ are Lorentz scalar functions. It is easy to prove that
\begin{eqnarray}
\slashed{v}\chi_{0P}(p)=\chi_{0P}(p),
\end{eqnarray}
\noindent hence we have~\cite{Guo96}
\begin{eqnarray}
\chi_{0P}(p)=\phi_{0P}(p)u_{\Lambda_{Q}}(v,s),
\end{eqnarray}
\noindent where $\phi_{0P}(p)$ is a scalar function.

However, $\slashed{v}\chi_{1P}(p)$ and $\slashed{v}\chi_{2P}(p)$ do not equal $\chi_{1P}(p)$ and $\chi_{2P}(p)$ respectively, so we define
\begin{eqnarray}
\chi^{+}_{1P}(p)=\frac{1+\slashed{v}}{2}\chi_{1P}(p),\nonumber\\
\chi^{-}_{1P}(p)=\frac{1-\slashed{v}}{2}\chi_{1P}(p),\nonumber\\
\chi^{+}_{2P}(p)=\frac{1+\slashed{v}}{2}\chi_{2P}(p),\nonumber\\
\chi^{-}_{2P}(p)=\frac{1-\slashed{v}}{2}\chi_{2P}(p).
\end{eqnarray}

It is easy to see that
\begin{eqnarray}
\slashed{v}\chi^{+}_{1,2P}(p)=\chi^{+}_{1,2P}(p),\\
\slashed{v}\chi^{-}_{1,2P}(p)=-\chi^{-}_{1,2P}(p).
\end{eqnarray}

Hence, like $\chi_{0P}(p)$, $\chi^{+}_{1,2P}(p)$ can be expressed as
\begin{eqnarray}
\chi^{+}_{1,2P}(p)=\phi^{+}_{1,2P}(p)u_{\Lambda_{Q}}(v,s),
\end{eqnarray}
\noindent where $\phi^{+}_{1P}(p)$ and $\phi^{+}_{2P}(p)$
are  scalar functions.

By using Eq.~(26) and (31), we can see that
\begin{eqnarray}
\chi^{-}_{1,2P}(p)=\slashed{p}_{t}\phi^{-}_{1,2P}(p)u_{\Lambda_{Q}}(v,s),
\end{eqnarray}
\noindent where again $\phi^{-}_{1P}(p)$ and $\phi^{-}_{2P}(p)$ are scalar functions. Then we obtain equations for all the scalar functions:
\begin{eqnarray}
\phi_{0P}(p)=\frac{-i}{(p_{l}+m_{D}+E_{0}+i\varepsilon)(p^{2}_{l}-W^{2}_{P}+i\varepsilon)}\int\frac{d^{4}q}{(2\pi)^{4}}[V_{1}+(p_{l}+q_{l})V_{2}]\phi_{0P}(q),
\end{eqnarray}
\begin{eqnarray}
\phi^{+}_{1P}(p)&=&\frac{-i}{(p_{l}+m_{D}+E_{0}+i\varepsilon)(p^{2}_{l}-W^{2}_{P}+i\varepsilon)}\int\frac{d^{4}q}{(2\pi)^{4}}[V_{3}+(p_{l}+q_{l})V_{4}]\phi_{0P}(q)\nonumber\\
&&+\frac{-i}{(p_{l}+m_{D}+E_{0}+i\varepsilon)(p^{2}_{l}-W^{2}_{P}+i\varepsilon)}\int\frac{d^{4}q}{(2\pi)^{4}}[V_{1}+(p_{l}+q_{l})V_{2}]\phi^{+}_{1P}(q)\nonumber\\
&&+\frac{-i(-2E_{1}+p^{2}_{t})}{2(p_{l}+m_{D}+E_{0}+i\varepsilon)^{2}(p^{2}_{l}-W^{2}_{P}+i\varepsilon)}\int\frac{d^{4}q}{(2\pi)^{4}}[V_{1}+(p_{l}+q_{l})V_{2}]\phi_{0P}(q),
\end{eqnarray}
\begin{eqnarray}
\phi^{-}_{1P}(p)&=&\frac{-i}{2(p_{l}+m_{D}+E_{0}+i\varepsilon)(p^{2}_{l}-W^{2}_{P}+i\varepsilon)}\int\frac{d^{4}q}{(2\pi)^{4}}[V_{1}+(p_{l}+q_{l})V_{2}]\phi_{0P}(q),
\end{eqnarray}
\begin{eqnarray}
\phi^{+}_{2P}(p)&=&\frac{-i}{(p_{l}+m_{D}+E_{0}+i\varepsilon)(p^{2}_{l}-W^{2}_{P}+i\varepsilon)}\int\frac{d^{4}q}{(2\pi)^{4}}[V_{1}+(p_{l}+q_{l})V_{2}]\phi^{+}_{2P}(q)\nonumber\\
&&+\frac{-i}{(p_{l}+m_{D}+E_{0}+i\varepsilon)(p^{2}_{l}-W^{2}_{P}+i\varepsilon)}\int\frac{d^{4}q}{(2\pi)^{4}}[V_{3}+(p_{l}+q_{l})V_{4}]\phi^{+}_{1P}(q)\nonumber\\
&&+\frac{-i}{(p_{l}+m_{D}+E_{0}+i\varepsilon)(p^{2}_{l}-W^{2}_{P}+i\varepsilon)}\int\frac{d^{4}q}{(2\pi)^{4}}(p_{t}\cdot q_{t}-q^{2}_{t})V_{4}\phi^{-}_{1P}(q)\nonumber\\
&&+\frac{-i}{(p_{l}+m_{D}+E_{0}+i\varepsilon)(p^{2}_{l}-W^{2}_{P}+i\varepsilon)}\int\frac{d^{4}q}{(2\pi)^{4}}[V_{5}+(p_{l}+q_{l})V_{6}]\phi_{0P}(q)\nonumber\\
&&+\frac{-i(-2E_{1}+p^{2}_{t})}{2(p_{l}+m_{D}+E_{0}+i\varepsilon)^{2}(p^{2}_{l}-W^{2}_{P}+i\varepsilon)}\int\frac{d^{4}q}{(2\pi)^{4}}[V_{1}+(p_{l}+q_{l})V_{2}]\phi^{+}_{1P}(q)\nonumber\\
&&+\frac{-i}{2(p_{l}+m_{D}+E_{0}+i\varepsilon)(p^{2}_{l}-W^{2}_{P}+i\varepsilon)}\int\frac{d^{4}q}{(2\pi)^{4}}[V_{1}+(p_{l}+q_{l})V_{2}]p_{t}\cdot q_{t}\phi^{-}_{1P}(q)\nonumber\\
&&+\frac{-i(-2E_{1}+p^{2}_{t})}{2(p_{l}+m_{D}+E_{0}+i\varepsilon)^{2}(p^{2}_{l}-W^{2}_{P}+i\varepsilon)}\int\frac{d^{4}q}{(2\pi)^{4}}[V_{3}+(p_{l}+q_{l})V_{4}]\phi_{0P}(q)\nonumber\\
&&+\Big[\frac{ip^{2}_{t}}{4(p_{l}+m_{D}+E_{0}+i\varepsilon)(p^{2}_{l}-W^{2}_{P}+i\varepsilon)}+\frac{iE_{2}}{(p_{l}+m_{D}+E_{0}+i\varepsilon)^{2}(p^{2}_{l}-W^{2}_{P}+i\varepsilon)}\nonumber\\
&&~~~~~~~~+\frac{-i(2E_{1}-p^{2}_{t})^{2}}{4(p_{l}+m_{D}+E_{0}+i\varepsilon)^{3}(p^{2}_{l}-W^{2}_{P}+i\varepsilon)}\Big]\int\frac{d^{4}q}{(2\pi)^{4}}[V_{1}+(p_{l}+q_{l})V_{2}]\phi_{0P}(q)\nonumber\\
&&+\frac{-i}{2(p_{l}+m_{D}+E_{0}+i\varepsilon)(p^{2}_{l}-W^{2}_{P}+i\varepsilon)}\int\frac{d^{4}q}{(2\pi)^{4}}[p_{t}\cdot q_{t}-p^{2}_{t}]V_{4}\phi_{0P}(q),
\end{eqnarray}
\begin{eqnarray}
\phi^{-}_{2P}(p)&=&\frac{i}{2(p^{2}_{l}-W^{2}_{P}+i\varepsilon)}\int\frac{d^{4}q}{(2\pi)^{4}}[V_{1}+(p_{l}+q_{l})V_{2}]\frac{p_{t}\cdot q_{t}}{p_{t}\cdot p_{t}}\phi^{-}_{1P}(q)\nonumber\\
&&+\frac{-i}{2(p_{l}+m_{D}+E_{0}+i\varepsilon)(p^{2}_{l}-W^{2}_{P}+i\varepsilon)}\int\frac{d^{4}q}{(2\pi)^{4}}[V_{1}+(p_{l}+q_{l})V_{2}]\phi^{+}_{1P}(q)\nonumber\\
&&+\frac{i}{2(p^{2}_{l}-W^{2}_{P}+i\varepsilon)}\int\frac{d^{4}q}{(2\pi)^{4}}\bigg(1+\frac{p_{t}\cdot q_{t}}{p_{t}\cdot p_{t}}\bigg)V_{4}\phi_{0P}(q)\nonumber\\
&&+\frac{-i}{2(p_{l}+m_{D}+E_{0}+i\varepsilon)^{2}(p^{2}_{l}-W^{2}_{P}+i\varepsilon)}\int\frac{d^{4}q}{(2\pi)^{4}}[V_{3}+(p_{l}+q_{l})V_{4}]\phi_{0P}(q)\nonumber\\
&&+\frac{i}{4(p^{2}_{l}-W^{2}_{P}+i\varepsilon)}\int\frac{d^{4}q}{(2\pi)^{4}}[V_{1}+(p_{l}+q_{l})V_{2}]\phi_{0P}(q)\nonumber\\
&&+\frac{i(2E_{1}-p^{2}_{t})}{4(p_{l}+m_{D}+E_{0}+i\varepsilon)^{2}(p^{2}_{l}-W^{2}_{P}+i\varepsilon)}\int\frac{d^{4}q}{(2\pi)^{4}}[V_{1}+(p_{l}+q_{l})V_{2}]\phi_{0P}(q).
\end{eqnarray}

From Eqs.~(34) and (36) it is easy to see that~\cite{Guo00}
\begin{eqnarray}
\phi^{-}_{1P}(p)=\frac{1}{2}\phi_{0P}(p).
\end{eqnarray}

The numerical solutions for $\phi_{0P}(p)$ and $\phi^{+}_{1P}(p)$ can be obtained by discretizing the integration region into $n$ pieces (with $n$ sufficiently large). In this way, the integral equations become matrix equations and the BS scalar functions $\phi_{0P}(p)$ and $\phi^{+}_{1P}(p)$ become $n$ dimensional vectors. Thus $\phi_{0P}(p)$ is the solution of the eigenvalue equation $(A-I)\phi_{0}=0$, where $A$ is an $n\times n$ matrix corresponding to the right hand side of Eq.~(34). In order to have a unique solution for the ground state, the rank of $(A-I)$ should be $n-1$. From Eq.~(35), $\phi^{+}_{1P}(p)$ is the solution of $(A-I)\phi_{1}=B$, where $B$ is an $n$ dimensional vector corresponding to the first and third integral terms on the right hand side of Eq.~(35). In order to have solutions for $\phi^{+}_{1P}(p)$, the rank of the augmented matrix $(A-I,B)$ should be equal to that of $(A-I)$, i.e., $B$ can be expressed as linear combination of the $n-1$ linearly independent columns in $(A-I)$. This is difficult to guarantee if $B\neq 0$, since the way to divide $(A-I)$ into $n$ columns is arbitrary. Therefore, following Ref.~\cite{Guo00} we demand the following condition in order to have solutions for $\phi^{+}_{1P}(p)$
\begin{eqnarray}
\int\frac{d^{4}q}{(2\pi)^{4}}\Bigg[\bigg(V_{3}+(p_{l}+q_{l})V_{4}\bigg)+\frac{-E_{1}+\frac{p^{2}_{t}}{2}}{p_{l}+m_{D}+E_{0}+i\varepsilon}\bigg(V_{1}+(p_{l}+q_{l})V_{2}\bigg)\Bigg]\phi_{0P}(q)=0.
\end{eqnarray}
The simplest forms for $V_{3}$ and $V_{4}$ satisfiying Eq.~(40) are
\begin{eqnarray}
V_{3}=\frac{E_{1}-\frac{p^{2}_{t}}{2}}{p_{l}+m_{D}+E_{0}+i\varepsilon}V_{1},
\end{eqnarray}
\begin{eqnarray}
V_{4}=\frac{E_{1}-\frac{p^{2}_{t}}{2}}{p_{l}+m_{D}+E_{0}+i\varepsilon}V_{2}.
\end{eqnarray}
With Eq.~(40), $\phi^{+}_{1P}(p)$ satisfies the same eigenvalue equation as $\phi_{0P}(p)$. Therefore, we have
\begin{eqnarray}
\phi^{+}_{1P}(p)=\sigma\phi_{0P}(p),
\end{eqnarray}
\noindent where $\sigma$ is a constant of proportionality with mass dimension. It was determined to be zero by applying Luke's theorem~\cite{Luk90} at the zero-recoil point in HQET~\cite{Guo00}. Therefore, $\phi^{+}_{1P}(p)$ does not contribute.

As to the forms of $V_{5}$ and $V_{6}$, we simply assume that
\begin{eqnarray}
V_{5}=\delta^{2}V_{1},\nonumber\\
V_{6}=\delta^{2}V_{2},
\end{eqnarray}
\noindent where $\delta$ is a parameter which is of order $\Lambda_{QCD}$.

From Eq.~(38) we can see that $\phi^{-}_{2P}(p)$ is only related to $\phi_{0P}(p)$ and $\phi^{+}_{1P}(p)$, so our goal is to solve the BS equations of $\phi_{0P}(p)$ and $\phi^{+}_{2P}(p)$ numerically.

In the covariant instantaneous approximation, $\tilde{V_{i}}= V_{i}|_{p_{l}=q_{l}},i=1,2$~\cite{Guo96}-\cite{Jin92}. In the meson case, one has~\cite{Jin92}
\begin{eqnarray}
\tilde{V}_{1}|_{meson}=\frac{8\pi\kappa'}{[(p_{t}-q_{t})^{2}+\mu^{2}]^{2}}-(2\pi)^{3}\delta^{3}(p_{t}-q_{t})\int\frac{d^{3}k}{(2\pi)^{3}}\frac{8\pi\kappa'}{(k^{2}+\mu^{2})^{2}},
\end{eqnarray}
\begin{eqnarray}
\tilde{V}_{2}|_{meson}=-\frac{16\pi}{3}\frac{\alpha_{seff}}{(p_{t}-q_{t})^{2}+\mu^{2}},
\end{eqnarray}
\noindent where $\kappa'$ and $\alpha_{seff}$ are coupling parameters related to the scalar confinement and the one-gluon-exchange terms, respectively. The parameter $\mu$ is introduced to avoid the infrared divergence in numerical calculations, and in the end we will take the limit $\mu\rightarrow0$.
In the baryon case, since scalar confinement term is still due to scalar interaction, the form of $\tilde{V_{1}}$ need not be changed. Only the parameter $\kappa'$ in the meson case has to be replaced by~$\kappa$ which describes the scalar confinement interaction between the heavy quark and the diquark. However, the diquark is not a pointlike object, there should be a form factor in $\tilde{V_{2}}$, $F(Q^{2})(Q=p_{2}-p'_{2})$, to describe the structure of the diquark (see FIG.1)~\cite{Ans87},
\begin{eqnarray}
F(Q^{2})=\frac{\alpha_{seff}Q^{2}_{0}}{Q^{2}+Q^{2}_{0}},
\end{eqnarray}
\noindent where $Q^{2}_{0}$ is a parameter which freezes $F(Q^{2})$ when $Q^{2}$ is very small. In the high energy region the form factor is proportional to $\frac{1}{Q^{2}}$ which is consistent with perturbative QCD calculations~\cite{Lep80}. By analyzing the electromagnetic form factor for the proton, it was found that $Q^{2}_{0}=3.2 GeV^{2}$ can lead to consistent results with the experimental data~\cite{Ans87}. Based on the above analysis , the form of kernel for BS equation in the baryon case is taken as~\cite{Guo96}
\begin{equation}
\tilde{V}_{1}=\frac{8\pi\kappa}{[(p_{t}-q_{t})^{2}+\mu^{2}]^{2}}-(2\pi)^{3}\delta^{3}(p_{t}-q_{t})\int\frac{d^{3}k}{(2\pi)^{3}}\frac{8\pi\kappa}{(k^{2}+\mu^{2})^{2}},
\end{equation}
\begin{eqnarray}
\tilde{V}_{2}=-\frac{16\pi}{3}\frac{\alpha^{2}_{seff}Q^{2}_{0}}{[(p_{t}-q_{t})^{2}+\mu^{2}][(p_{t}-q_{t})^{2}+Q^{2}_{0}]}.
\end{eqnarray}

From the BS equation solutions in the meson case it was found that the values $m_{b}=5.02 GeV$ and $m_{c}=1.58 GeV$ give predictions which are in good agreement with experiments~\cite{Jin92}. Hence in the baryon case we take
\begin{eqnarray}
m_{D}+E_{0}+\frac{1}{m_{b}}E_{1}+\frac{1}{m^{2}_{b}}E_{2}=0.62GeV.
\end{eqnarray}

The dimension of $\kappa$ is three and that of $\kappa'$ is two. This extra dimension in $\kappa$ should be caused by nonperturbative diagrams which include the frozen form factor $F(Q^{2})$ at low momentum region. Since $\Lambda_{QCD}$ is the only parameter which is related to confinement, we expect that
\begin{eqnarray}
\kappa=\delta\kappa',
\end{eqnarray}

\noindent where $\delta$ is of order $\Lambda_{QCD}$. It is noted that the two proportionality parameters in Eqs.~(44) and (51) could be different in general although they are both of order $\Lambda_{QCD}$. However, in order to simplify our model we assume these two parameters are the same. The value of $\kappa$ was determined to be in the region between $0.02GeV^{3}$ and $0.08GeV^{3}$ in Ref.~\cite{Guo07}. Since $\kappa'$ is about $0.2GeV^{2}$~\cite{Jin92}, we take $\delta$ to be in the range from $0.1GeV$ to $0.4GeV$.

One notes that $E_{1}\sim \Lambda_{QCD}E_{0}$, $E_{2}\sim \Lambda_{QCD}E_{1}$, hence we further assume that $E_{1}=\delta E_{0}$, $E_{2}=\delta E_{1}=\delta^{2}E_{0}$.

\begin{table}
\caption{Values of $\alpha_{seff}$ for different $m_{D}$ and $\kappa$.}
\begin{tabular}{c c c c c c}\hline
 & $\kappa(GeV^{3})$ & 0.02& 0.04 & 0.06 & 0.08 \\
$m_{D}=650MeV$ &$\alpha_{seff}$ & 0.66 & 0.7  & 0.73 & 0.75 \\
$m_{D}=700MeV$ &$\alpha_{seff}$ & 0.69 & 0.73 & 0.76 & 0.78 \\
$m_{D}=750MeV$ &$\alpha_{seff}$ & 0.74 & 0.77 & 0.79 & 0.8 \\
$m_{D}=800MeV$ &$\alpha_{seff}$ & 0.77 & 0.8  & 0.82 & 0.83\\
\hline
\end{tabular}
\end{table}

In general, $\phi_{0P}(p)$ can be the function of $p_{l}$ and $p_{t}$. Defining $\tilde{\phi}_{0P}(p_{t})\equiv\int(dp_{l}/2\pi)\phi_{0P}$, one gets immediately the BS equation for $\tilde{\phi}_{0P}(p_{t})$:
\begin{eqnarray}
\tilde{\phi}_{0P}(p_{t})=-\frac{1}{2(-W_{p}+m_{D}+E_{0})W_{p}}\int\frac{d^{3}q_{t}}{(2\pi)^{3}}(\tilde{V}_{1}-2W_{p}\tilde{V}_{2})\tilde{\phi}_{0P}(q_{t}),
\end{eqnarray}

\noindent where we have used the residue theorem and selected the upper contour in the $p_{l}$  plane which has a singular point $-W_{p}+i\varepsilon$ (see FIG.2).

\begin{center}
\begin{figure}[htbp]
\includegraphics[width=10cm]{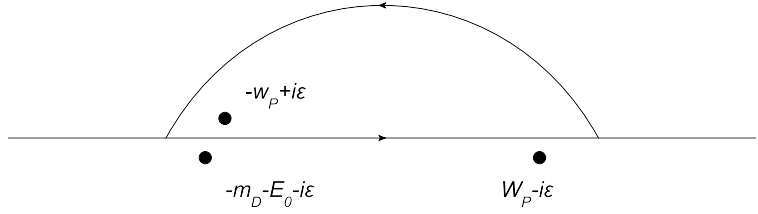}
\caption{~Three singular points, $-m_{D}-E_{0}-i\varepsilon$, $-W_{p}+i\varepsilon$ and $W_{p}-i\varepsilon$, in the $p_{l}$ plane.}
\end{figure}
\end{center}

Substituting $\tilde{V}_{1}$ and $\tilde{V}_{2}$ into Eq.~(52) we have~\cite{Guo96}
\begin{eqnarray}
(-W_{p}+m_{D}+E_{0})\tilde{\phi}_{0P}(p_{t})&=&-\frac{1}{2W_{p}}\Bigg\{\int\frac{q^{2}_{t}dq_{t}}{4\pi^{2}}\frac{16\pi\kappa}{(p^{2}_{t}+q^{2}_{t}+u^{2})^{2}-4p^{2}_{t}q^{2}_{t}}\tilde{\phi}_{0P}(q_{t})+\frac{32\pi\alpha^{2}_{seff}Q^{2}_{0}W_{p}}{3(Q_{0}-u^{2})}\nonumber\\
&&\times\int\frac{q^{2}_{t}dq_{t}}{4\pi^{2}}\frac{1}{2p_{t}q_{t}}\bigg[ln\frac{(p_{t}+q_{t})^{2}+u^{2}}{(p_{t}-q_{t})^{2}+u^{2}}-ln\frac{(p_{t}+q_{t})^{2}+Q^{2}_{0}}{(p_{t}-q_{t})^{2}+Q^{2}_{0}}\bigg]\tilde{\phi}_{0P}(q_{t})\Bigg\}\nonumber\\
&&+\frac{1}{2W_{p}}\int\frac{q^{2}_{t}dq_{t}}{4\pi^{2}}\frac{16\pi\kappa}{(p^{2}_{t}+q^{2}_{t}+u^{2})^{2}-4p^{2}_{t}q^{2}_{t}}\tilde{\phi}_{0P}(p_{t}).
\end{eqnarray}

This is a eigenvalue equation of $\tilde{\phi}_{0P}(p_{t})$. Solving this equation we obtain the values of the model parameter $\alpha_{seff}$ corresponding to different values of $\kappa$ and $m_{D}$ (see TABLE 1).

With the numerical result of $\tilde{\phi}_{0P}(p_{t})$, we can solve out $\tilde{\phi}^{+}_{2P}(p_{t})$ numerically from Eq.~(37).

The normalization equation for the BS equation takes the following form~\cite{Lur68}:
\begin{eqnarray}
\frac{i}{(2\pi)^{4}}\int d^{4}qd^{4}q'\bar{\chi}_{k}(q')\frac{\partial}{\partial k_{0}}[I(q',q,k)+K(q',q,k)]\chi_{k}(q)=2k_{0},
\end{eqnarray}

\noindent where $I(q',q,P)=\delta^{(4)}(q'-q)[S_{F}(p_{1})]^{-1}[S_{D}(-p_{2})]^{-1}$,
in which $p_{1}=\lambda_{1}P+q', p_{2}=-\lambda_{2}P+q'$, and $k_{0}$ represents the energy of $\Lambda_{b}$, which is the mass of $\Lambda_{b}$ in the rest frame of $\Lambda_{b}$. With this condition we can give the plots of normalized $\tilde{\phi}_{0P}(p_{t})$ and $\tilde{\phi}^{+}_{2P}(p_{t})$ in FIGs. 3 and 4, respectively.

\begin{center}
\begin{figure}[htbp]
\includegraphics[width=8cm]{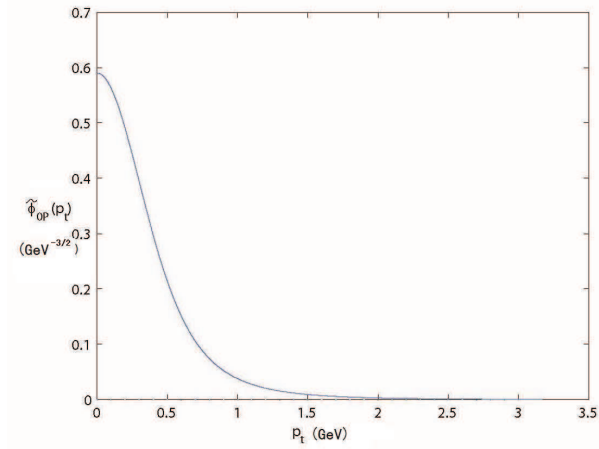}
\caption{ Plot of~$\tilde{\phi}_{0P}(p_{t})$ with $m_{D}=650MeV$,~$\kappa=0.04$.}
\end{figure}
\end{center}

\begin{center}
\begin{figure}[htbp]
\includegraphics[width=8cm]{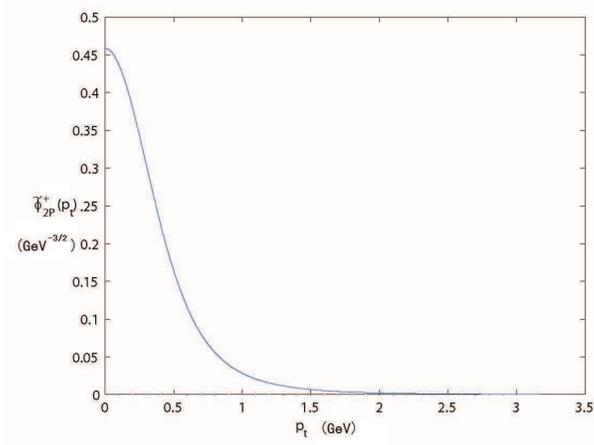}
\caption{ Plot of $\tilde{\phi}^{+}_{2P}(p_{t})$ with ~$m_{D}=650MeV$,~$\kappa=0.04$.}
\end{figure}
\end{center}

\section*{VI. Numerical results of~$\epsilon_{b}$ }

With the numerical results of the BS amplitudes solved out above, we can calculate the value of $\epsilon_{b}$. The left hand side of Eq.~(5) can be expressed as the overlap integral of the BS amplitude as the following:
\begin{eqnarray}
\langle\Lambda_{b}(v,s)|\bar{b}\gamma^{\lambda}\gamma_{5}b|\Lambda_{b}(v,s)\rangle=\int\frac{d^{4}p}{(2\pi)^{4}}\bar{\chi}_{P}(p)\gamma^{\lambda}\gamma_{5}\chi_{P}(p)S^{-1}_{D}(p_{2}).
\end{eqnarray}

Hence we have
\begin{eqnarray}
(1+\epsilon_{b})\bar{u}_{\Lambda_{b}}(v,s)\gamma^{\lambda}\gamma_{5}u_{\Lambda_{b}}(v,s)=\int\frac{d^{4}p}{(2\pi)^{4}}\bar{\chi}_{P}(p)\gamma^{\lambda}\gamma_{5}\chi_{P}(p)S^{-1}_{D}(p_{2}).
\end{eqnarray}

Expanding both sides of Eq.~(56) to order $1/m^{2}_{Q}$ , we have
\begin{eqnarray}
&&\bigg(1+\epsilon_{0b}+\frac{1}{m_{b}}\epsilon_{1b}+\frac{1}{m^{2}_{b}}\epsilon_{2b}\bigg)\bar{u}_{\Lambda_{b}}(v,s)\gamma^{\lambda}\gamma_{5}u_{\Lambda_{b}}(v,s)\nonumber\\
&=&\int\frac{d^{4}p}{(2\pi)^{4}}\bar{u}_{\Lambda_{b}}(v,s)\bigg[\phi_{0P}(p)+\frac{1}{m_{b}}\big(\phi^{+}_{1P}(p)+\phi^{-}_{1P}(p)\slashed{p}_{t}\big)+\frac{1}{m^{2}_{b}}\big(\phi^{+}_{2P}(p)+\phi^{-}_{2P}(p)\slashed{p}_{t}\big)\bigg]\nonumber\\
&&\times\gamma^{\lambda}\gamma_{5}\bigg[\phi_{0P}(p)+\frac{1}{m_{b}}(\phi^{+}_{1P}(p)+\phi^{-}_{1P}(p)\slashed{p}_{t})+\frac{1}{m^{2}_{b}}\Big(\phi^{+}_{2P}(p)\phi^{-}_{2P}(p)\slashed{p}_{t}\Big)\bigg]u_{\Lambda_{b}}(v,s)S^{-1}_{D}(p_{2}),
\end{eqnarray}
\noindent where we have expanded $\epsilon_{b}$ in Eq.~(5) as
\begin{eqnarray}
\epsilon_{b}=\epsilon_{0b}+\frac{1}{m_{b}}\epsilon_{1b}+\frac{1}{m^{2}_{b}}\epsilon_{2b}.
\end{eqnarray}

By comparing the two sides of Eq.~(57) at each order in the $1/m_{Q}$ expansion, we have the following equations:

\noindent to leading order in the $1/m_{Q}$ expansion:
\begin{eqnarray}
(1+\epsilon_{0b})\bar{u}_{\Lambda_{b}}(v,s)\gamma^{\lambda}\gamma_{5}u_{\Lambda_{b}}(v,s)=\int\frac{d^{4}p}{(2\pi)^{4}}\bar{u}_{\Lambda_{b}}(v,s)\phi_{0P}(p)\gamma^{\lambda}\gamma_{5}\phi_{0P}(p)u_{\Lambda_{b}}(v,s)S^{-1}_{D}(p_{2}),
\end{eqnarray}

\noindent to first order in the $1/m_{Q}$ expansion:
\begin{eqnarray}
\epsilon_{1b}\bar{u}_{\Lambda_{b}}(v,s)\gamma^{\lambda}\gamma_{5}u_{\Lambda_{b}}(v,s)&=&\int\frac{d^{4}p}{(2\pi)^{4}}\bar{u}_{\Lambda_{b}}(v,s)[\phi_{0P}(p)\gamma^{\lambda}\gamma_{5}\phi^{+}_{1P}(p)+\phi_{0P}(p)\gamma^{\lambda}\gamma_{5}\slashed{p}_{t}\phi^{-}_{1P}(p)\nonumber\\
&&+\phi^{+}_{1P}(p)\gamma^{\lambda}\gamma_{5}\phi_{0P}(p)+\phi^{-}_{1P}(p)\slashed{p}_{t}\gamma^{\lambda}\gamma_{5}\phi_{0P}(p)]u_{\Lambda_{b}}(v,s)S^{-1}_{D}(p_{2}),
\end{eqnarray}

\noindent to second order in the $1/m_{Q}$ expansion:
\begin{eqnarray}
\epsilon_{2b}\bar{u}_{\Lambda_{b}}(v,s)\gamma^{\lambda}\gamma_{5}u_{\Lambda_{b}}(v,s)&=&\int\frac{d^{4}p}{(2\pi)^{4}}\bar{u}_{\Lambda_{b}}(v,s)[\phi_{0P}(p)\gamma^{\lambda}\gamma_{5}\phi^{+}_{2P}(p)+\phi_{0P}(p)\gamma^{\lambda}\gamma_{5}\slashed{p}_{t}\phi^{-}_{2P}(p)+\phi^{+}_{1P}(p)\gamma^{\lambda}\gamma_{5}\phi^{+}_{1P}(p)\nonumber\\
&&+\phi^{+}_{1P}(p)\gamma^{\lambda}\gamma_{5}\slashed{p}_{t}\phi^{-}_{1P}(p)+\phi^{-}_{1P}(p)\slashed{p}_{t}\gamma^{\lambda}\gamma_{5}\phi^{+}_{1P}(p)+\phi^{-}_{1P}(p)\slashed{p}_{t}\gamma^{\lambda}\gamma_{5}\slashed{p}_{t}\phi^{-}_{1P}(p)\nonumber\\
&&+\phi^{+}_{2P}(p)\gamma^{\lambda}\gamma_{5}\phi_{0P}(p)+\phi^{-}_{2P}(p)\slashed{p}_{t}\gamma^{\lambda}\gamma_{5}\phi_{0P}(p)]u_{\Lambda_{b}}(v,s)S^{-1}_{D}(p_{2}).
\end{eqnarray}

Getting rid of the spinors, $\epsilon_{b}$ at each order in the $1/m_{Q}$ expansion can be expressed as the following:
\begin{eqnarray}
(1+\epsilon_{0b})=\int\frac{d^{4}p}{(2\pi)^{4}}\phi_{0P}(p)\phi_{0P}(p)S^{-1}_{D}(p_{2}),
\end{eqnarray}
\begin{eqnarray}
\epsilon_{1b}=2\int\frac{d^{4}p}{(2\pi)^{4}}\phi_{0P}(p)\phi^{+}_{1P}(p)S^{-1}_{D}(p_{2}),
\end{eqnarray}
\begin{eqnarray}
\epsilon_{2b}&=&2\int\frac{d^{4}p}{(2\pi)^{4}}\phi_{0P}(p)\phi^{+}_{2P}(p)(p^{2}_{l}-W^{2}_{P})-\frac{1}{3}\int\frac{d^{4}p}{(2\pi)^{4}}\phi^{-}_{1P}(p)\phi^{-}_{1P}(p)p^{2}_{t}(p^{2}_{l}-W^{2}_{P}).
\end{eqnarray}

From Eq.~(62) one can see immediately that $\epsilon_{0b}=0$ because the Isgur-Wise function is normalized to 1. From Eq.~(63) we have $\epsilon_{1b}=0$ since $\phi^{+}_{1P}(p)=0$. This is consistent with the demand of the current conservation~\cite{Luk90}.

Substituting Eqs.~(34),~(37), and (39) into Eq.~(64), using $\tilde{\phi}(p_{t})=\int\frac{dp_{l}}{2\pi}\phi(p)$ and the residue theorem, we obtain the value of $\epsilon_{2b}$ with the obtained numerical results of $\tilde{\phi}_{0P}(p_{t})$ and $\tilde{\phi}^{+}_{2P}(p_{t})$. The results are listed in TABLE II.

We can see from TABLE II that the value of $\epsilon_{2b}$ depends on our model parameters, $m_{D}$ and $\kappa$. In the ranges of the parameters, $\epsilon_{2b}$ is always negative. The absolute value of $\epsilon_{2b}$ increases with the increase of $m_{D}$ and $\kappa$. When $m_{D}=650MeV$, $\sqrt{|\epsilon_{2b}|}$ changes from $0.33GeV$ to $0.56GeV$. When $m_{D}=800MeV$, $\sqrt{|\epsilon_{2b}|}$ changes from $0.62GeV$ to $0.79GeV$. As expected, $\sqrt{|\epsilon_{2b}|}$ is of order $\Lambda_{QCD}$. This indicates that our results are reasonable.

\begin{table}
\caption{Values of $\epsilon_{2b}$ for different $m_{D}$ and $\kappa$.}
\begin{tabular}{c c c c c c}\hline
 & $\kappa(GeV^{3})$ & 0.02& 0.04 & 0.06 & 0.08 \\
$m_{D}=650MeV$ &~~$\epsilon_{2b}(GeV^{2})$~~& -0.11 & -0.19 & -0.26 & -0.32 \\
$m_{D}=700MeV$ &~~$\epsilon_{2b}(GeV^{2})$~~& -0.16 & -0.25 & -0.32 & -0.41 \\
$m_{D}=750MeV$ &~~$\epsilon_{2b}(GeV^{2})$~~& -0.27 & -0.36 & -0.44 & -0.49 \\
$m_{D}=800MeV$ &~~$\epsilon_{2b}(GeV^{2})$~~& -0.39 & -0.47 & -0.56 & -0.62 \\
\hline
\end{tabular}
\end{table}

\section*{IV. Summary and discussion}

More and more data have been and will be collected at LHC. This makes more precise measurement of Cabibbo-Kobayashi-Maskawa matrix elements, especially the value of $V_{ub}$, become possible. $V_{ub}$ can be measured through the inclusive semileptonic decays of polarized $\Lambda_{b}$. Since the decay rate involves two parameters, $\mu^{2}_{\pi}$ and $\epsilon_{b}$, to second order in the $1/m_{b}$ expansion, the theoretical calculation of these two parameters is important to extract more precise $V_{ub}$ from experimental data. $\mu^{2}_{\pi}$ was calculated in the previous work. In this paper we focus on the theoretical calculation of $\epsilon_{b}$ in the BS equation approach. Since $\epsilon_{b}$ is only non-zero at second order in the $1/m_{b}$ expansion, it is necessary to establish the BS equation for $\Lambda_{b}$ to this order.

We regard the heavy baryon $\Lambda_{b}$ as composed of a heavy $b$ quark and a light scalar diquark based on the fact that the light degrees of freedom in $\Lambda_{b}$ have fixed spin and isospin quantum numbers. In this picture, we establish the BS equation for $\Lambda_{b}$ to second order in the $1/m_{b}$ expansion. Then we solve the BS equation numerically by applying the kernel which include the scalar confinement and the one-gluon-exchange terms. To second order in the $1/m_{b}$ expansion, we obtain numerical results for BS scalar functions. Expressing $\epsilon_{b}$ as the overlap integral of the BS amplitude for $\Lambda_{b}$ we obtain the numerical values for $\epsilon_{b}$. It is found that $\epsilon_{b}$ is only non-zero at $1/m^{2}_{b}$ as expected. This indicates that our result is consistent with the requirements from the heavy quark symmetry and the current conservation. At $1/m^{2}_{b}$, $\epsilon_{b}$ can be expressed as $\epsilon_{2b}/m^{2}_{b}$. We found that in the ranges of our model parameters, $\epsilon_{2b}$ is always negative, and varies form $-0.11GeV^{2}$ to $-0.62GeV^{2}$. We find that $\sqrt{|\epsilon_{2b}|}$ is of order $\Lambda_{QCD}$ as expected.

There are some uncertainties in our model. Compared with the heavy meson case, heavy baryons are much more complicated. To order $1/m_{b}$ there are two parameters in our model, $m_{D}$ and $\kappa$~\cite{Guo00}. When we expand to second order in the $1/m_{b}$ expansion, more uncertainties are involved, including the forms for the kernel $V_{5}$ and $V_{6}$ and the values of binding energies at orders $1/m_{b}~(E_{1})$ and $1/m^{2}_{b}~(E_{2})$. To simplify our model we assume that $V_{5}=\delta^{2}V_{1}$, $V_{6}=\delta^{2}V_{2}$, $E_{2}=\delta E_{1}=\delta^{2}E_{0}$, with the same proportionality parameter $\delta$ which is of order $\Lambda_{QCD}$. Our model to first order in the $1/m_{Q}$ expansion has been tested in $\Lambda_{b}\rightarrow\Lambda_{c}$ semileptonic decay and $\Lambda_{b}\rightarrow\Lambda_{c}\pi$ nonleptonic decay~\cite{Guo96}\cite{Guo00}\cite{Ber12}. Further tests are needed in other physical processes and for the model to $1/m^{2}_{Q}$.

\begin{acknowledgements}
This work was partially supported by Natural Science Foundation of China under Contract No. 10975018, No. 11175020, No. 11275025 and the Fundamental Research Funds for the Central Universities of China.
\end{acknowledgements}

\end{document}